\documentclass
{iopart}
\usepackage{graphicx}
\usepackage[T1]{fontenc}
\usepackage[latin1]{inputenc}
\usepackage[thinspace,thinqspace]{SIunits}					

\newcommand{\YRS}{YbRh$_2$Si$_2$}
\newcommand{\YIS}{YbIr$_2$Si$_2$}

\newcommand{\TN}{\ensuremath{T_{\mathrm{N}}}}									
\newcommand{\mT}{\milli\tesla}														
\newcommand{\dd}{\mathrm{d}} 															
\newcommand{\mK}{\milli\kelvin}
\newcommand{\RH}{\ensuremath{R_{\mathrm H}}}												
\newcommand{\TRH}{\ensuremath{\tilde{R}_{\mathrm H}}}												
\newcommand{\TRHO}{\ensuremath{\tilde{R}_{\mathrm H}^0}}												
\newcommand{\TRHI}{\ensuremath{\tilde{R}_{\mathrm H}^{\infty}}}												
\newcommand{\RHI}{\ensuremath{\RH^{\infty}}}												
\newcommand{\RHO}{\ensuremath{\RH^{0}}}												
\newcommand{\rhoH}{\ensuremath{\rho_{\text H}}}										

\newcommand{\sven}[1]{%
{#1}}

\newcommand{\svenzwei}[1]{%
{#1}}

\begin{document}

\title{Discontinuous Hall coefficient at the quantum critical point in \YRS}

\author{Sven Friedemann$^1$, Niels Oeschler$^1$, Steffen Wirth$^1$, Cornelius Krellner$^1$, Christoph Geibel$^1$, Frank Steglich$^1$, Silke Paschen$^2$, Stefan Kirchner$^{3,4}$ and Qimiao Si$^3$}

\address{$^1$  Max Planck Institute for Chemical Physics of Solids, Nöthnitzer Str. 40, 01187 Dresden, Germany}
\address{$^{2}$Institute of Solid State Physics, TU Vienna, Wiedner Hauptstrasse 8-10, 1040~Vienna, Austria}
\address{$^{3}$Department of Physics and Astronomy, Rice University, Houston, TX 77005, USA}
\address{$^{4}$Max Planck Institute for the Physics of Complex Systems, Nöthnitzer Str. 38, 01187 Dresden, Germany}

\ead{Sven.Friedemann@cpfs.mpg.de}

\begin{abstract}
\YRS\ is a model system for quantum criticality. Particularly, Hall effect measurements helped identify the unconventional nature of its quantum critical point. 
Here, we present a high-resolution study of the Hall effect and magnetoresistivity on samples of different quality. We find a robust crossover on top of a sample dependent linear background contribution. Our detailed analysis provides a complete characterization of the crossover in terms of its position, width, and height. Importantly, we find in the extrapolation to zero temperature a 
discontinuity of the Hall coefficient occurring at the quantum critical point for all samples. 
Particularly, the height of the jump in the Hall coefficient remains finite in the limit of zero temperature.
Hence, our data  solidify the conclusion of a collapsing Fermi surface. 
Finally, we contrast our results to the 
smooth Hall-effect evolution seen in Chromium, the prototype
system for a spin-density-wave quantum critical point.
\end{abstract}

\submitto{JPCM}

\section{Introduction}

A quantum critical point (QCP) marks a continuous phase transition at zero temperature. A non-thermal parameter such as pressure or magnetic field is used to tune a material from one state to another. Experimentally, QCPs are often accessed by suppressing a finite temperature phase transition.
At zero temperature quantum effects become dominant. Surprisingly, the quantum fluctuations between the two incompatible ground states lead to a rich variety of phenomena even at finite temperatures. In fact, quantum critical phenomena are believed to be relevant for a large variety of correlated materials  such as quantum magnets, heavy-fermion materials, and even high-temperature superconductors \cite{Focus2008}.

Heavy-fermion materials play a major role in the study of quantum critical phenomena \cite{Gegenwart2008}. These intermetallic compounds appear to be model systems because two competing interactions leading to two different ground states may be shifted relative to each other allowing to tune the material from one ground state to another. Both, the Kondo interaction and the Ruderman-Kittel-Kasuya-Yosida (RKKY) interaction arise from the interplay between conduction electrons and magnetic $f$ electrons, the latter being provided by rare earth elements. The Kondo effect marks the screening of $f$ electrons by conduction electrons leading to composite quasiparticles with largely enhanced effective masses. These quasiparticles arise in the Kondo-screened spin-singlet ground state.
Whereas the Kondo effect favors a paramagnetic Landau-Fermi-liquid (LFL) ground state the RKKY interaction mediates a magnetic exchange between $f$-electrons and, hence, favors long range magnetic order. 
The different dependences of the Kondo effect and the RKKY interaction on the coupling strength of conduction electrons and $f$ electrons provides the basis for tuning heavy-fermion materials to quantum criticality in a controlled fashion: Pressure for instance increases the hybridization of conduction electrons and $f$ moments, hence, increasing the Kondo coupling strength.
The Kondo effect dominates in the strong coupling regime  whereas the RKKY interaction dominates in the weak coupling regime. At an intermediate coupling strength, both balance each other such that a magnetic transition arising at weak coupling due to the RKKY interaction is suppressed to zero temperature giving rise to the QCP.

The conventional description of a QCP is based on the Ginzburg-Landau concept of order-parameter fluctuations. 
It generalizes its counterpart for classical critical points 
by incorporating the quantum fluctuations.
For the metallic heavy-fermion systems the magnetism is treated itinerantly as a spin-density-wave (SDW) \cite{Hertz1976,Millis1993,Moriya1995}. Within this picture, the composite quasiparticles are assumed to stay intact at the QCP: They form both the magnetically ordered ground state on one side of the QCP as well as the paramagnetic heavy Fermi liquid on the other side. In addition, the fact that the QCP corresponds to a Gaussian fixed point implies that the fluctuations violate energy over temperature, $E/T$, scaling \cite{Sachdev1999}. In disagreement with this fundamental property of a conventional QCP the magnetic fluctuations in CeCu$_{5.9}$Au$_{0.1}$ were observed to obey an $E/T$ scaling \cite{Schroeder2000}. This stimulated new theoretical approaches.

In unconventional theories of the QCP additional quantum modes are assumed to become critical. These additional critical modes may lead to an interacting fixed point for which the fluctuations satisfy $E/T$ scaling in contrast to the SDW scenario, hence, explaining the observations in CeCu$_{5.9}$Au$_{0.1}$. For the case of the heavy-fermion materials these additional modes are associated with the breakdown of the Kondo effect \cite{Si2001,Coleman2001,Senthil2003}. As a consequence, the composite quasiparticles are expected to disintegrate at the QCP leaving the $f$ electrons decoupled from the conduction electron sea on the weak coupling side. Consequently, the Fermi surface is expected to change from "large" with the $f$ states incorporated on the strong coupling side to "small" with the $f$ states decoupled from the conduction electron sea on the weak coupling side. As such a Fermi surface collapse is not expected in the conventional scenario measurements of the Fermi surface evolution were suggested to unveil the nature of a particular heavy-fermion QCP \cite{Si2001,Coleman2001}.  In fact, 
Hall effect measurements gave indications of a discontinuous evolution across the QCP in \YRS\ \cite{%
Paschen2004}. 
Here, a crossover of the Hall coefficient is observed at finite temperatures. As the temperature is lowered this crossover sharpens. With the width of the crossover vanishing in the extrapolation to zero temperature an abrupt jump of the Hall coefficient at the QCP is suggested. Consequently, these measurements indicate a reconstruction of the Fermi surface at the QCP.
However, this Hall effect study was discussed controversially particularly in view of sample dependences in the low temperature Hall coefficient \cite{Friedemann2008}: The Hall coefficient is extremely sensitive to small changes in the relative scattering rates of the two dominating bands which almost compensate each other \cite{Friedemann2010c}. These changes in the scattering rates seem to originate from tiny variations in the chemical composition as they only affect samples from different batches.

In order to identify the influence of the sample dependences on the behavior at the QCP a high-resolution Hall effect study on different samples of \YRS\ was carried out \cite{Friedemann2010b}. This study revealed the crossover in the Hall coefficient to be robust against sample dependences. Moreover, the results gave indications for the electronic fluctuations to obey $E/T$ scaling. Consequently, \YRS\ seems to be the first material in which both the fundamental signatures of a Kondo breakdown QCP are seen, \textit{i.e.}, the Fermi surface reconstruction and $E/T$ scaling. Given this key role of \YRS\ for the understanding of quantum critical phenomena it is important to carefully scrutinize the picture of a Fermi surface reconstruction. 

Here, we present extended measurements and a  detailed analysis of the magnetotransport properties across the QCP surveying different samples. 
Our detailed analysis of the data at very low temperatures allows to establish a proper extrapolation of the characteristics of the Hall coefficient to zero temperature. 
We show that these characteristics are distinct
from the theoretical predictions for a SDW QCP and the experimental observations
in the canonical SDW QCP system, Chromium.

\section{Experimental Setup}
Pronounced non-Fermi-liquid behavior of \YRS\ such as a linear temperature dependence of the resistivity and a divergent specific heat were interpreted in terms of its proximity to a QCP. In zero field, \YRS\ orders antiferromagnetically below the Néel temperature $\TN = \unit{70}\mK$ \cite{Trovarelli2000}. The application of a small magnetic field of $B_{\text c2}=\unit{60}\mT$ applied perpendicular to the crystallographic $c$ direction or of $B_{\text c1}=\unit{660}\mT$ applied along the crystallographic $c$ axis suppresses the magnetic order to zero temperature thus accessing the field-induced QCP \cite{Gegenwart2002}. 
For fields exceeding the critical field a paramagnetic LFL ground state is observed.

Single crystals of \YRS\ were grown in Indium flux as described earlier \cite{Trovarelli2000}. By optimizing the growth procedure, the quality of the crystals was advanced. Two different samples were chosen: Sample 1 possessing a residual resistivity ratio  $\text{RRR}=70$  was taken from Ref.~\cite{Paschen2004}. Sample 2 with $\text{RRR}=120$ was taken from the highest quality batch available.

\begin{figure}%
	\setlength{\unitlength}{.5\textwidth}
	\begin{picture}(1,.71)
  		\put(0,0){\includegraphics[width=\unitlength]{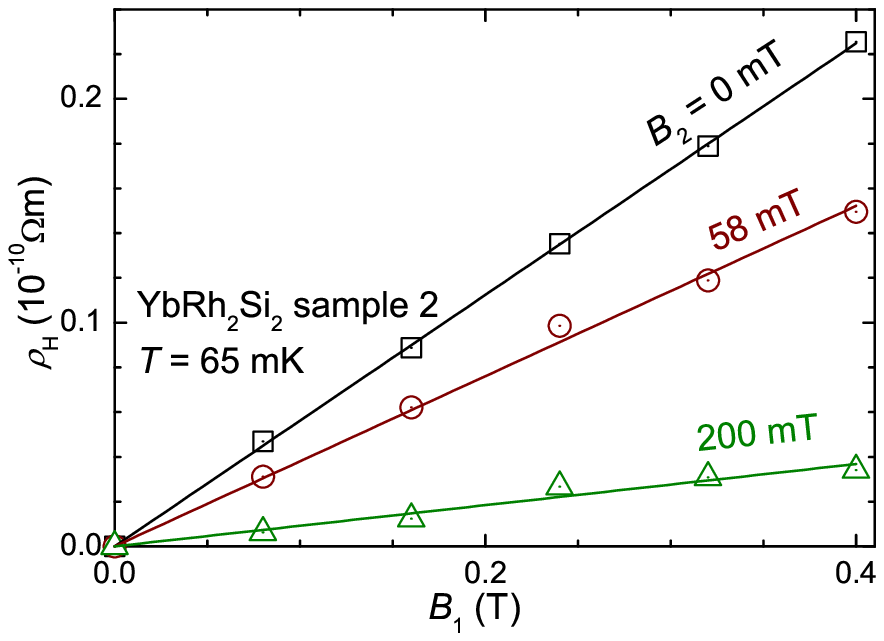}}
  		\put(.17,.425){\includegraphics[width=.5\unitlength]{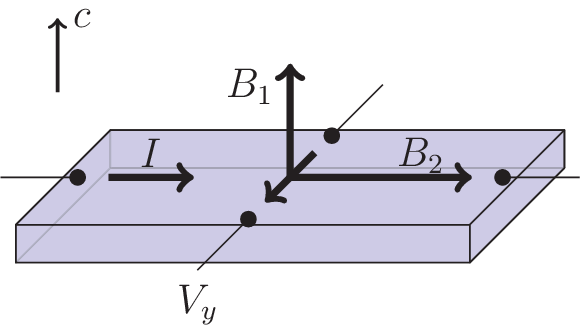}}
  	\end{picture}
	\caption{Exemplary Hall resistivity isotherms for sample 2. Solid lines mark linear fits to the data with the slope of these lines corresponding to the linear-response Hall coefficient \RH. Inset sketches the crossed-field Hall effect setup.}
	\label{fig:RhoH}%
\end{figure}

We utilize three different magnetotransport techniques to study the Fermi surface evolution of \YRS. First, the crossed-field Hall effect geometry is used to measure the linear-response Hall coefficient. Here, two different magnetic fields allow to disentangle the two responses to the magnetic field  (cf.~inset in Fig.~\ref{fig:RhoH}): A small field $B_1$ oriented along the $c$ direction and perpendicular to the current $I$ is used to induce a transverse Hall voltage. A second field $B_2$ within the $ab$ plane and parallel to the current is used to tune the sample across the QCP.
The tuning effect of $B_1$ is negligible thanks to the magnetic anisotropy seen for instance in the ratio of the critical fields.
The Hall resistivity \rhoH\ is calculated as the antisymmetric part of the transversal voltage $V_y$ with respect to $B_1$ in order to separate magnetoresistance contributions, 
$\rhoH (B_1,B_2) = t/(2I) \times [V_y(B_1,B_2) - V_y(- B_1, B_2)]$.
The Hall resistivity obeys a linear field dependence as displayed in Fig.~\ref{fig:RhoH}. The initial-slope Hall coefficient \RH\ was subsequently numerically extracted as the slope of linear fits to the Hall resistivity 
\begin{equation}
\RH (B_2) \equiv \lim_{B_1\to 0} \frac{\rhoH(B_1,B_2)}{B_1}
\label{eq:RH}
\end{equation}
 at fixed tuning field $B_2$ for Hall fields $B_1\leq\unit{0.4}\tesla$. Consequently, the decrease of the slope in $\rhoH(B_1)$  as the field $B_2$ is increased, as seen in Fig.~\ref{fig:RhoH}, corresponds to a decrease of the Hall coefficient. Second, the results are corroborated by measurements in standard single-field setup. Here, only the magnetic field $B_1$ (parallel to $c$) is used to perform both tasks. By analyzing the differential Hall coefficient
\begin{equation}
\TRH (B_1)= \frac{\partial \rhoH(B_1,0)}{\partial B_1}
\label{eq:TRH}
\end{equation} 
we subsequently disentangle these two effects. 
Third, longitudinal magnetoresistivity with current flowing in the $ab$ plane was monitored over an extended temperature range and provides increased statistics.

For the electrical transport measurements, rectangular platelets with a thickness $t$ between \unit{25}\micro\metre\ and \unit{80}\micro\metre\ were used. Spot welded gold wires provided contacts with resistances well below \unit{1}\ohm. An alternating current $I$ up to \unit{100}\micro\ampere\ flowed within the $ab$ plane. Voltages arising from magnetoresistance and Hall effect were amplified with low-temperature transformers and subsequently monitored via standard lock-in technique. 
For \YRS, the anomalous contributions to the Hall coefficient \cite{Fert1987}
are negligible at low temperatures \cite{Paschen2005,Friedemann2010b}.

\section{Results and Discussion}
\begin{figure}%
	\includegraphics[width=.5\textwidth]{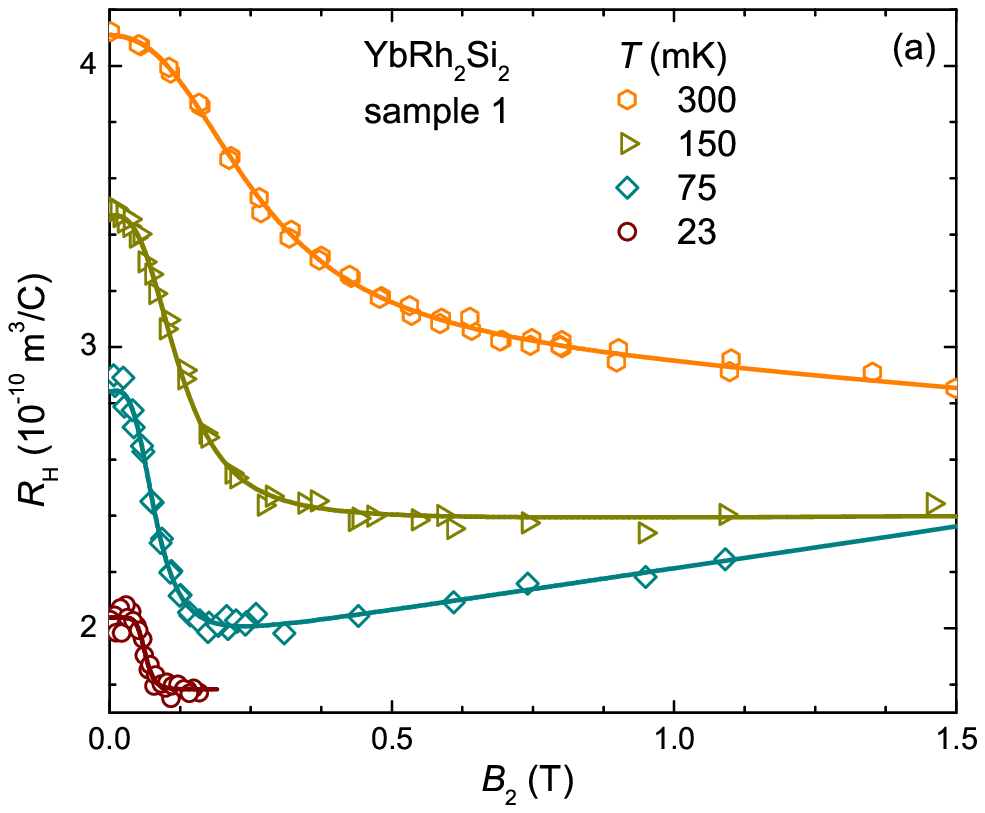}%
	\includegraphics[width=.5\textwidth]{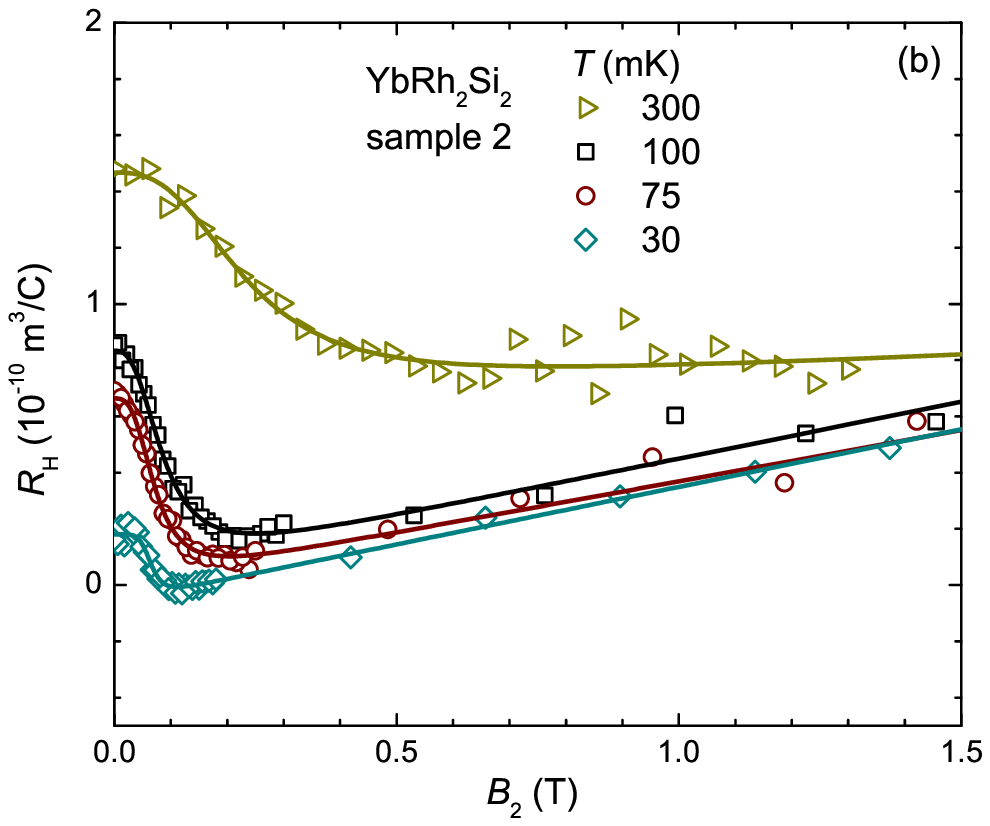}%
	\caption{Field dependence of the initial-slope Hall coefficient, $\RH(B_2)$, for sample 1 (a) and 2 (b). Sample 1 was taken from Ref.~\cite{Paschen2004}. Solid lines represent fits of eq.~\ref{eq:Crossover} to the data.}%
	\label{fig:RHvsB}%
\end{figure}

Low temperature isotherms of the linear-response Hall coefficient as a function of $B_2$ are displayed in Fig.~\ref{fig:RHvsB}. 
Both samples show a similar qualitative behavior consisting of two features: At small fields, the Hall coefficient decreases in a pronounced crossover and at higher fields and low temperatures, this crossover is succeeded by a linear increase. The crossover was already earlier recognized as being part of the intrinsic quantum critical behavior \cite{Paschen2004} whereas the linear increase is only apparent in the extended field and temperature range studied here. Both the crossover as well as the linear high-field behavior obey temperature dependences: The crossover in $\RH(B_2)$ is shifted towards smaller fields approaching the QCP and becomes sharper as the temperature is lowered. Hence, the linear behavior is revealed over an increasing field range as the temperature is lowered. This suggests that the linear behavior represents a background contribution which presumably originates from Zeeman splitting. The slope of this background contribution is reduced above \unit{100}\mK\ and becomes slightly negative for sample 1 at \unit{300}\mK. Despite the robust characteristics of the curves, the absolute values of the Hall coefficient differ for the two samples. This is in accordance to the above mentioned sample dependences.

\begin{figure}%
	\includegraphics[width=.85\textwidth]{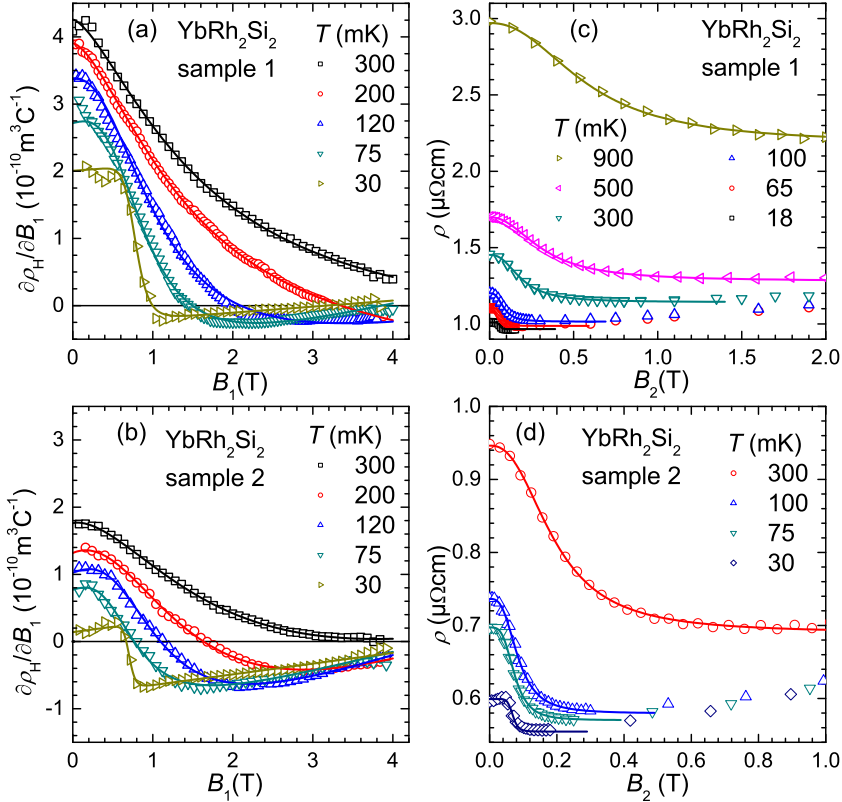}%
	\caption{Differential Hall coefficient (left) and magnetoresistivity (right) isotherms of sample 1 (upper panels) and sample 2 (lower panels).  Solid lines mark fits of eq.~\ref{eq:Crossover} to the data. In case of the magnetoresistivity the linear term ($m=0$) was omitted, as it better describes the data.}%
	\label{fig:dRhoHdB1}%
\end{figure}

Differential Hall coefficient and magnetoresistivity are depicted in Fig.~\ref{fig:dRhoHdB1} for both samples. Like in the linear-response Hall coefficient, we can identify a crossover and a linear background contribution. Again, the crossover shifts to smaller fields and becomes sharper as the temperature is lowered. Also, a decrease of the crossover height is seen in both differential Hall coefficient and magnetoresistivity. Nevertheless, the absolute height of the crossover in the differential Hall coefficient is larger compared to the linear-response Hall coefficient. 
Finally, the signatures are seen in both samples with the absolute values varying according to the sample dependences of the Hall coefficient and the different residual resistivity of the samples, respectively.

The presence of the crossover in both samples and in various transport properties suggests that this signature of the QCP is not affected by  sample dependences. Rather, the sample dependences seem to be exclusively related to the background contribution \cite{Friedemann2010c}.
In order to check this we shall quantitatively analyze the crossover for its position and width. In addition, our high-resolution study over an extended temperature range reveals the height of the crossover to become smaller with decreasing temperature. This motivates a careful analysis of the limiting parameters of the Hall crossover in order to check if the change of the Hall coefficient is retained at the QCP, \textit{i.e.}, in the extrapolation to zero temperature.
In order to scrutinize this, we analyze the Hall coefficient by fitting the empirical crossover function
\begin{equation}
\RH(B_2) = \RH^{\infty}+m B_2-\frac{\RH^{\infty}-\RH^0}{1+(B_2/B_0)^p} 
\label{eq:Crossover}
\end{equation}
to the data.
Here, $\RH^0$ and $\RH^{\infty}$ parametrize the zero-field and high-field value, respectively. The position of the crossover is represented by $B_0$ and its sharpness is determined by $p$. The superposed linear term $m B_2$ is added to reflect the  background behavior.
Analog fitting procedures lead to the zero-field and high-field values of the single-field Hall experiment (\TRHO\ and \TRHI) and of the magnetoresistivity experiment ($\rho^0$ and $\rho^{\infty}$), respectively.
Eq.~\ref{eq:Crossover} allows us to analyze the characteristics of the crossover separately from the linear background. 

\begin{figure}%
	\includegraphics[width=.5\textwidth]{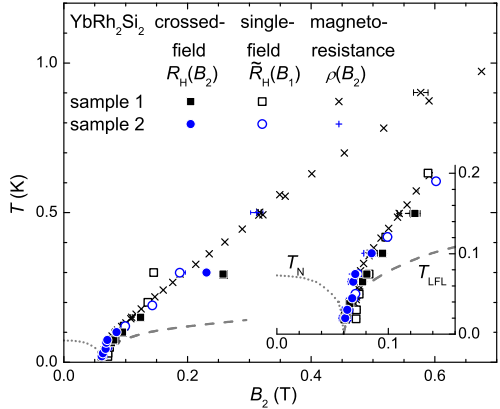}%
	\caption{Position of the crossover fields from Hall effect $\RH(B_2)$ and $\TRH(B_1)$ and magnetoresistivity $\rho(B_2)$ in the magnetic field-temperature phase diagram (cf.~Figs.~\ref{fig:RHvsB} and \ref{fig:dRhoHdB1}). Single-field results are scaled by $B_{\text c 2}/B_{\text c 1}=1/11$ accounting for the magnetic anisotropy of \YRS \cite{Gegenwart2002}. 
	\svenzwei{Inset magnifies the low-temperature range.}
	Dotted and dashed line represent the boundary of the magnetically ordered state and the boundary of the regime with LFL behavior as deduced from resistivity measurements \cite{Gegenwart2002}.}%
	\label{fig:PDRH_LargeScale}%
\end{figure}

First, we examine the position of the crossover.  
Fig.~\ref{fig:PDRH_LargeScale} depicts the crossover field $B_0$ extracted from the three experiments and for both samples in the low temperature-magnetic field phase diagram. We note that in the case of the single-field Hall experiment, the magnetic anisotropy ratio $B_{\text c 2}/B_{\text c 1}=1/11$ is used to convert the results to an equivalent $B_2$ scale. For both samples and the complete set of experiments $B_0$ is seen to shift to lower fields as the temperature is decreased. In the limit of zero temperature it extrapolates to \unit{60}\mT, \textit{i.e.}, to the QCP \svenzwei{as illustrated in the inset of Fig.~\ref{fig:PDRH_LargeScale}}. 
This 
emphasizes the robust assignment of the crossover to the quantum criticality.

\begin{figure}%
	\includegraphics[width=.5\textwidth]{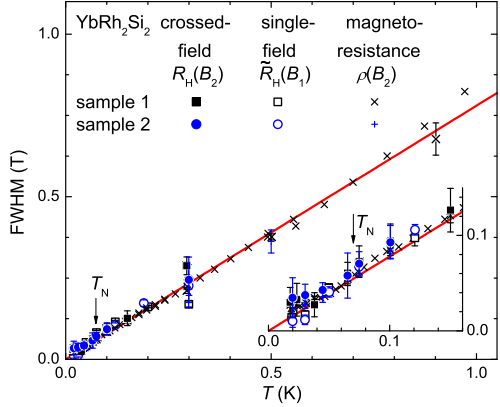}%
	\caption{Full width at half maximum (FWHM) of the crossovers in the Hall coefficient and magnetoresistivity. Single-field results are scaled by $B_{\text c 2}/B_{\text c 1}=1/11$. The FWHM was extracted from the derivative  of eq.~(\ref{eq:Crossover}) fitted to the linear response Hall coefficient $\RH(B_2)$, differential Hall coefficient $\TRH(B_1)$, and magnetoresistivity $\rho(B_2)$ (cf.~Figs.~\ref{fig:RHvsB} and \ref{fig:dRhoHdB1}). Solid line represents a linear fit to all data sets. Within the error this line intersects the origin. \svenzwei{Inset magnifies data at lowest temperatures.} Arrow indicates the Néel temperature.}%
	\label{fig:FWHM_LargeScale}%
\end{figure}

Second, in order to quantify the width of the crossover the derivative of the fitted function is analyzed. The crossover in $\RH(B_2)$ corresponds to a (negative) peak in the derivative $\dd \RH / \dd B_2$ and analogously in $\dd \TRH/\dd B_1$ and $\dd \rho/\dd B_2$. We examine the full width at half maximum (FWHM) of this peak as plotted in Fig.~\ref{fig:FWHM_LargeScale}. The analysis reveals, that also the width of the crossover is unaffected by the sample dependences: Within the experimental resolution the data of the two samples match. Moreover, the different experiments yield identical results within experimental accuracy. The FWHM decreases as the temperature is lowered. Importantly, $\text{FWHM}(T)$ \svenzwei{is best described by} a proportionality to temperature extrapolating to zero for $T=0$. \svenzwei{In fact, no signature is seen at the Néel temperature in contrast to the height of the crossover discussed below. At temperatures below \unit{30}{mK} the crossed-field data show a slight trend towards saturation which is however absent in the single-field data. This might indicate that the classical fluctuations play a role at these lowest temperatures where the Hall crossover significantly interferes with the classical transition at \TN. The difference between the crossed-field and single field results might arise from differently strong classical fluctuations. They are presumably strongly enhanced for the crossed-field orientation with the field in the magnetic easy plane as here the magnetization is by one order of magnitude larger compared to the single-field orientation. Finally, we note that within experimental accuracy the data are well described over the complete range $\unit{18}\mK \leq T \leq \unit{1}\kelvin$ by a proportionality to temperature. This implies a vanishing width at $T\to 0$  meaning} that the crossover becomes infinitely sharp in the zero temperature limit with such an abrupt change suggesting a sudden reconstruction of the Fermi surface at the QCP. 

\begin{figure*}%
	\includegraphics[width=\textwidth]{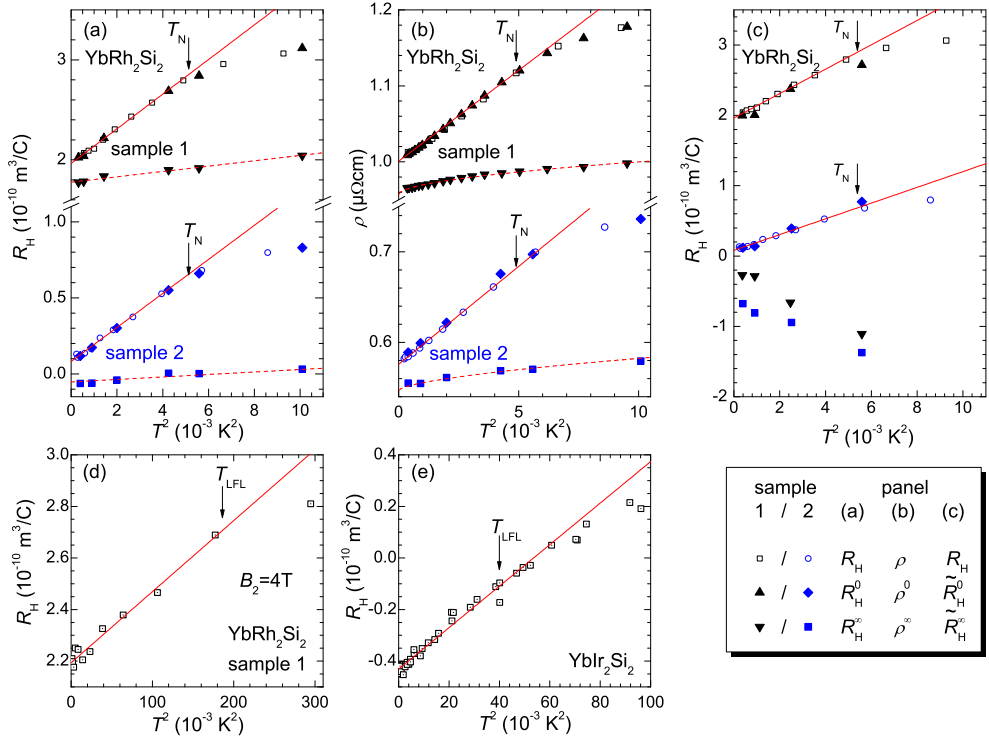}%
	\caption{Low-temperature behavior of the Hall coefficient and resistivity. (a) Initial-slope Hall coefficient is plotted against temperature squared.  The zero-field Hall coefficient $\RHO$ and high-field Hall coefficient $\RHI$ extracted from the fits to the field-dependence of the linear-response Hall coefficient are included (cf.\ Fig.~\ref{fig:RHvsB}). (b) Resistivity of both \YRS\ samples  together with the zero-field value $\rho^0$ and $\rho^{\infty}$ extracted from the fits to the magnetoresistivity (Figs.~\ref{fig:dRhoHdB1}(c) and (d)). (c) Analog parameter of the differential Hall coefficient: \TRHO\ and \TRHI\ are plotted against $T^2$ together with the initial-slope Hall coefficient for both samples.  Arrows in (a),  (b), and (c) mark the Néel temperature. (d) Initial-slope Hall coefficient of \YRS\ against $T^2$ measured at a constant tuning field $B_2 = \unit{4}\tesla$, \textit{i.e.}, in the field induced Fermi-liquid state. (e) Hall coefficient of \YIS\ against $T^2$. Arrows in (d) and (e) mark the Fermi-liquid temperature below which the resistivity obeys a quadratic temperature dependence \cite{Gegenwart2008,Hossain2005}. Solid lines in (a) - (e) represent fits of a quadratic temperature dependence to the data. Dashed lines are a guide to the eye.}%
	\label{fig:RHOuIvsTsq}%
\end{figure*}

Finally, for a full characterization of the crossover in the Hall coefficient we examine its height. The limiting values of the crossover in the linear-response Hall coefficient, \textit{i.e.}, $\RH^0$ and $\RH^{\infty}$ are plotted in Fig.~\ref{fig:RHOuIvsTsq}(a). For a proper evaluation we first analyze the low temperature behavior of the measured initial-slope Hall coefficient in zero tuning field, \textit{i.e.}, $\RH(T,B_2=0)$. The representation against $T^2$ in Fig.~\ref{fig:RHOuIvsTsq}(a) reveals a quadratic temperature dependence setting in just below the Néel temperature as previously observed for the electrical resistivity \cite{Gegenwart2002}. Such a quadratic temperature dependence of the Hall coefficient is likely to arise from finite temperature corrections within a Fermi-liquid description. An enhancement like for the corresponding term in the resistivity might render this term observable in heavy-fermion materials. In fact, we observe a quadratic temperature dependence of \RH\  not only in the magnetically ordered phase but also in the field induced Fermi-liquid state of \YRS, and in the paramagnetic ground state of the heavy-fermion material \YIS\  as shown in Figs.~\ref{fig:RHOuIvsTsq} (d) and (e). The quadratic form of $\RH(T)$ is limited to the regime where also resistivity obeys a quadratic temperature dependence.

As a next step we compare the zero-field Hall coefficient \RHO\ extracted from the analysis of the crossover with the truly measured zero-field ($B_2=0$) Hall coefficient. This comparison is non-trivial as \RHO\ was allowed to vary during the fitting procedure. Nevertheless, we find a perfect agreement proving the consistency of our analysis. Consequently, we can utilize the power law found for $\RH(T)$ to extrapolate \RHO\ to zero temperature.
This is highlighted by the solid lines in Fig.~\ref{fig:RHOuIvsTsq}(a). The evolution of \RHO\ is to be compared with that of the high-field value $\RH^{\infty}$. A distinct power law can not be established for $\RH^{\infty}$ due to the limited statistics. Nevertheless, the fact that \RHI\ features only a small temperature dependence allows a proper extrapolation of this parameter as indicated by the dashed lines in Fig.~\ref{fig:RHOuIvsTsq}(a).  Importantly, the finite difference between \RHO\ and \RHI\ persists down to zero temperature for both samples. Complimentary conclusions can be derived for the parameters extracted from the analysis of the magnetoresistivity and differential Hall coefficient. In fact, $\rho^0$ and $\rho^{\infty}$ plotted in Fig.~\ref{fig:RHOuIvsTsq}(b) behave very similar to \RHO\ and \RHI. A quadratic form of $\rho(T)$ and $\rho^0(T)$ sets in just below \TN. Also, the high-field value $\rho^{\infty}$ resembles the behavior of \RHI. On this basis a proper extrapolation of both $\rho^0$ and $\rho^{\infty}$ yields  a persistent finite difference in the limit of zero temperature. For the case of the differential Hall coefficient the zero-field value \TRHO\ depicted in Fig.~\ref{fig:RHOuIvsTsq}(c) follows the very same quadratic form of the initial-slope Hall coefficient as \RHO, thus further proving the consistency of our analysis. The only difference between the crossed-field and single-field results affects the high-field values \RHI\ and \TRHI. Whereas \RHI\ is seen to increase with increasing temperature, \TRHI\ decreases. In addition, the absolute values of both differ, \TRHI\ appears to be reduced and  negative for both samples whereas \RHI\ is only negative for sample 2 at lowest temperatures. As a consequence the difference $\TRHO-\TRHI$ is larger than $\RHO-\RHI$ reflecting the larger stepheight seen in the single-field experiment (compare Fig.~\ref{fig:RHvsB} and \ref{fig:dRhoHdB1} (a) and (b)). For the extrapolation of the parameters of the single-field experiment we again take advantage of the quadratic form for \TRHO\ and the smooth evolution of \TRHI. As for the crossed-field results, the finite difference $\TRHO-\TRHI$ persists in the zero-temperature limit. In summary, the difference between the zero-field value and the high-field value of the crossover in all three quantities remains finite down to zero temperature for both samples. 
Together with the vanishing FWHM and the position of the crossover converging to the QCP this difference marks a \emph{finite} jump of the Hall coefficient. Consequently, the data strongly suggest, that the Fermi surface undergoes a severe reconstruction at the QCP. The difference between \RHO\ and \RHI\ is associated with the different Fermi surfaces of the adjacent different ground states (cf. Ref. \cite{Friedemann2010c}).

\sven{We can rule out that the discontinuity of \RH\ at the QCP originates from a change in scattering rates only. In fact, a slight shift in the balance of scattering rates for the two dominant bands is invoked to explain the strong sample dependences of the background contribution in \RH. The sample dependences \svenzwei{in $\RH(T)$ arise for} different samples with minute differences in composition and different values of disorder. However, the (extrapolated) jump in \RH\ across the QCP is a property associated with criticality; moreover, it occurs as a function of a \textit{continuous} variation of the control parameter -- the magnetic field \sven{-- at a \textit{continuous} phase transition. In fact, magnetostriction measurements show that the phase transition remains continuous down to at least 15\,mK \cite{Kuechler2004}.} Without a jump of Fermi surface, a change in \RH\ resulting from a variation of scattering rates \textit{per se} \sven{across a continuous phase transition}  must be continuous with respect to the control parameter and cannot have a jump across the QCP.}

These strong indications for a Fermi surface reconstruction at the QCP in \YRS\ provide the basis for the scaling analysis of the Hall crossover width. The linear-in-temperature form of the FWHM was found to point towards an $E/T$ scaling of the critical fluctuations \cite{Friedemann2010b}. Our magnetoresistivity measurements establish the proportionality of the FWHM with temperature to persist up to \unit{1}\kelvin, \textit{i.e.}, over almost two decades. Consequently, these results underpin the conclusion of an unconventional QCP in \YRS.

\section{Comparison with spin-density-wave quantum critical point}

\begin{figure}%
	\includegraphics[width=.5\textwidth]{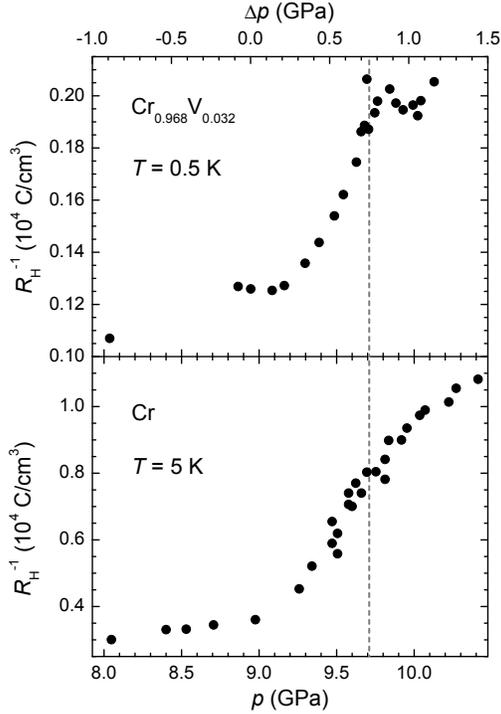}%
	\caption{Inverse of the Hall coefficient of ${\rm Cr_{0.968}V_{0.032}}$ (upper panel)
	and Cr (lower panel) as a function of pressure. In the upper panel, $\Delta p$
	is the external pressure for ${\rm Cr_{0.968}V_{0.032}}$, and an effective
	pressure when the V-concentration differs from $0.032$ \cite{Lee2004}. In the
	lower panel, the result is shown only in the vicinity of the critical 
	pressure \cite{Jaramillo2010} so that the two panels cover the same pressure extent. 
	The dashed lines label the critical pressure.}%
	\label{fig:Cr}%
\end{figure}

A canonical SDW QCP is realized in pure and V-doped Cr \cite{Yeh2002}. 
The evolution of the Hall coefficient across the pressured-driven
QCPs 
has been systematically studied in both cases \cite{Lee2004, Jaramillo2010},
and is shown in Fig.~\ref{fig:Cr} for the lowest measured temperatures.
These temperatures (5 K and 0.5 K for pure Cr and ${\rm Cr_{0.968}V_{0.032}}$)
are already small compared to the natural temperature scales of the system, 
yet the Hall coefficient is seen to be smoothly evolving as a function
of the control parameter. Such a smooth evolution is expected
theoretically \cite{Coleman2001,Norman2003,Bazaliy2004}. 
The Fermi surface of a SDW state is reconstructed from
that of the paramagnetic state through a band folding, which is more
pronounced for a system like Cr whose Fermi surfaces are nested. However,
when the SDW order parameter is adiabatically switched off, the folded 
Fermi surface is smoothly connected to the paramagnetic one. As a result, the
Hall coefficient does not show a jump provided the nesting is not 
perfect \cite{Bazaliy2004}.

In a magnetic-field driven QCP, the nonzero critical field
can give rise to a small discontinuity in the magnetotransport coefficients
even in the case of an SDW QCP \cite{Fenton2005}. Approaching the QCP, the energy scale
associated with the vanishing order parameter becomes smaller than the scale
associated with the Lorentz force, leading to a non-linearity in the 
system's response to the Lorentz force \sven{reflected in a breakdown of the weak-field magnetotransport.
The linear field dependence of the magnetoresistance in Ca$_3$Ru$_2$O$_7$ was taken as indication thereof \cite{Kikugawa2010}.
More significantly, the breakdown of weak-field magnetotransport may lead to a jump in the Hall coefficient which, however, does not appear to have been seen in any of the other quantum critical systems. 
This is not surprising as the jump in the Hall coefficient is likely to be very small.
In fact, for the case of the cuprates mean field theory predicts that the anomaly in the Hall response
at the critical doping level is non-observable small even though the SDW gap is large \cite{Lin2005}.
Disorder may smear the jump of the Hall coefficient into
a smooth crossover.}

As described above, in the pressure-driven SDW QCP of the 
V-doped and pure Cr \cite{Lee2004, Jaramillo2010}, the Hall coefficient (and the
resistivity) is smooth across the QCP with a crossover width that does not
track with the strength of disorder (Fig.~\ref{fig:Cr}) 
\sven{in contrast to the scenario of a breakdown of the weak-field limit.}
 For the field-driven
QCP of \YRS, the issue of non-linear response to the Lorentz force
is completely avoided by the cross-field Hall setup, in which the Lorentz
field $B_1$ can be vanishingly small while the tuning field $B_2$ is fixed
at its critical value. As seen in Fig.~1, the Hall resistivity is indeed linear 
in $B_1$ even at the critical $B_2$. Our conclusion is reinforced by
considerations of several other factors. Our measured width of the critical
crossover component is the same for both samples (1 and 2) that have different
amounts of disorder (Fig.~\ref{fig:FWHM_LargeScale}). The fact that the single-field
Hall and magnetoresistivity have the same crossover width as the cross-field
Hall effect implies that, even in our single-field measurements, the
critical Hall crossover does not originate from the above-noted 
non-linear effect. 
This last conclusion is \sven{not surprising},
given that at the critical $B_{\text c 1}$ field, $\omega_c \tau$ is of the order
of 0.01 and 0.002 for samples 1 and 2 respectively.

\section{Conclusion}

Our study of the Hall effect and magnetoresistivity on the model system \YRS\ provides a systematic characterization of the signatures of unconventional quantum criticality in this material. We analyze various magnetotransport properties  for samples of different quality. We find a robust crossover in the Hall coefficient and magnetoresistivity on top of a sample dependent background. The crossover sharpens towards a finite jump at the QCP in the extrapolation to zero temperature,
indicating a collapse of the Fermi surface
The relevance of such a Fermi surface reconstruction for the concept underlying our understanding of quantum-critical metals cannot be overstated. Given the continuous nature of the transition the collapse indicates that the Fermi surface is much more fragile than expected. Hence, it is necessary to introduce a new class of quantum phase transitions in metals.
We also argue that our results are very different from
the expectations for and observations in SDW QCPs.

%
%
%
%
%
%
%
%
%
%
%
%
%

\ack
Fruitful discussions with with P. Gegenwart, \sven{T. Rosenbaum}, A. J. Schofield and P. Wölfle as well as partial support by the DFG Research Group 960 ``Quantum Phase Transitions'' are acknowledged. {S.~P.\ acknowledges
funding from the European Research Council under the European Community's 
Seventh Framework Programme (FP7/2007-2013)/ERC grant agreement
n$^{\circ}$~227378.} S.~K.\ and Q.~S.\ were supported by the NSF \sven{Grant No.~DMR-1006985} and
the Welch Foundation Grant No. C-1411 and S.~F.\ was partially supported by the Alexander von Humboldt foundation.
\section*{References}

\providecommand{\newblock}{}

\end{document}